\documentclass[namedreferences,here]{kluwer}

\newcommand{\ltapprox}{\rlap{\raise.4ex\hbox{$<$}}\lower.5ex\hbox
{{$\sim$}}}
\newcommand{\gtapprox}{\rlap{\raise.4ex\hbox{$>$}}\lower.5ex\hbox
{{$\sim$}}}

\let\oldverbatim\verbatim
\renewcommand{\verbatim}{\expandafter\small\oldverbatim}
\usepackage{epsfig}

\runningtitle{CIRCUIT ESTIMATES OF MAGNETIC FIELD ENERGIES}

\runningauthor{M.S.\ WHEATLAND AND F.J.\ FARVIS}

\begin{document}

\begin{opening}

\title{TESTING CIRCUIT MODELS FOR THE ENERGIES OF CORONAL MAGNETIC FIELD 
CONFIGURATIONS}

\author{M.S. \surname{WHEATLAND}}
\author{F.J. \surname{FARVIS}}
\institute{School of Physics, University of Sydney, NSW 2006, Australia
 \protect\\(e-mail: wheat@physics.usyd.edu.au)}


\date{}


\begin{ao}
\end{ao}

\begin{abstract}
Circuit models involving bulk currents and inductances are often used to
estimate the energies of coronal magnetic field configurations, in 
particular configurations associated with solar flares. The accuracy of 
circuit models is tested by comparing calculated energies of linear 
force-free fields with specified boundary conditions with corresponding
circuit estimates. The circuit models are found to provide reasonable 
(order of magnitude) estimates for the energies of the non-potential 
components of the fields, and to reproduce observed functional dependences 
of the energies. However, substantial departure from the circuit estimates
is observed for large values of the force-free parameter, and this is
attributed to the influence of the non-potential component of the field 
on the path taken by the current. 
\end{abstract}

\end{opening}
\keywords{Sun: magnetic fields --- Sun: corona --- Sun: activity Sun: flares}

\section{Introduction}

\noindent
Solar flares are understood to involve the conversion of magnetic 
energy in the solar corona into the energy of particles and radiation. 
In particular flare energy is believed to derive from non-potential coronal
magnetic fields, i.e.\ that part of the magnetic field produced by 
currents flowing in the solar corona (e.g.\ Tandberg-Hanssen and 
Emslie, 1988). 

The origin of flare energy in coronal currents has suggested to 
many authors the use of a circuit analogy (e.g.\ Alfv\'{e}n and
Carlqvist 1967; Spicer, 1982; Somov 1992; Melrose, 1995; Melrose 1997; 
Longcope and Noonan 2000; Khodachenko and Zaitsev, 1998; Khodachenko,
Haerendel and Rucker, 2003).
It is well known that the energy of a system of $N$ distinct 
current-carrying circuits may be written (e.g.\ Jackson, 1998)
\begin{equation}\label{eq:general_E}
E=\frac{1}{2}\sum_{i=1}^{N}L_iI_i^2
  +\sum_{i=1}^{N}\sum_{j>i}M_{ij}I_iI_j,
\end{equation}
where
\begin{equation}\label{eq:general_L}
L_i=\frac{\mu_0}{4\pi I_i^2}\int_{C_i}d^3x_i\int_{C_i}d^3x_i^{\prime}
  \frac{{\bf J}({\bf x}_i)\cdot {\bf J}({\bf x}_i^{\prime})}
  {|{\bf x}_i-{\bf x}_i^{\prime}|}
\end{equation}
is the self-inductance and 
\begin{equation}\label{eq:general_M}
M_{ij}=\frac{\mu_0}{4\pi I_iI_j}\int_{C_i}d^3x_i\int_{C_j}d^3x_j^{\prime}
  \frac{{\bf J}({\bf x}_i)\cdot {\bf J}({\bf x}_j^{\prime})}
  {|{\bf x}_i-{\bf x}_j^{\prime}|}
\end{equation}
is the mutual inductance, and where ${\bf J}({\bf x})$ is
the current density at a point ${\bf x}$ in space, and $C_i$ is the
volume of the $i^{\rm th}$ circuit. Although 
Equations~(\ref{eq:general_E})---(\ref{eq:general_M}) are quite general,
their utility relies on having a small number of well-defined circuits
with simple geometries for which analytic expressions for $L_i$ and 
$M_{ij}$ may be evaluated, and in particular for which $L_i$ and $M_{ij}$
are independent of $I_i$ and $I_j$. This is the situation for currents 
flowing in a small number of ``wire'' loops, i.e.\ loops with a fixed
geometry.  Circuit models for coronal magnetic fields adopt 
wire loop configurations. For example, in the flare model of Melrose (1997) 
based on reconnection between current-carrying magnetic 
flux tubes, each flux tube is represented by fixed toroidal structures 
carrying a uniform axial current. The self-inductance of
such a structure is (e.g.\ Landau and Lifshitz, 1960)
\begin{equation}\label{eq:L}
L_i=\mu_0 a_i C_i, \quad \quad 
  C_i= \ln \frac{8a_i}{r_i}-\frac{7}{4},
\end{equation}
where $a_i$ is the major radius of the toroid and $r_i$ is the minor radius.
Melrose (1997) also adopts the following approximate expression for the 
mutual inductance between two loops whose centres are separated by a 
distance $d_{ij}$ and whose axes subtend an angle $\theta_{ij}$:
\begin{equation}\label{eq:M_don}
M_{ij}=\mu_0 \frac{8a_i^2a_j^2 (C_iC_j)^{1/2}\cos\theta_{ij}}
  {\left[(a_i+a_j)^2+d_{ij}^2\right]^{3/2}}.
\end{equation}
This formula is not exact but is based on interpolation between known 
results. An exact formula which holds for two axially-aligned line-current 
loops separated by a distance $d_{ij}$ is (Stratton, 1941)
\begin{equation}\label{eq:stratton}
M_{ij}=\mu_0(a_ia_j)^{1/2}\left[\left(\frac{2}{k_{ij}}-k_{ij}\right)
K(k_{ij}) -\frac{2}{k_{ij}}E(k_{ij})
\right],
\end{equation} 
where
\begin{equation}
\quad k_{ij}^2=\frac{4a_ia_j}{(a_i+a_j)^2+d_{ij}^2},
\end{equation}
and where $K(x)$ and $E(x)$ are complete elliptic integrals.

Real coronal currents do not flow along simple wire loops --- they are 
distributed in the coronal volume, and describe complicated paths. A 
common model for current-carrying coronal fields is the force-free 
model, in which the magnetic field ${\bf B}$ satisfies 
${\bf J}\times {\bf B}=0$, 
(as well as $\nabla\cdot {\bf B}=0$)
where ${\bf J}=\frac{1}{\mu_0}\nabla\times {\bf B}$. In other words the
current density is everywhere parallel to the magnetic field.
Unfortunately the force-free field 
equation is in general nonlinear and hence difficult to solve. A restricted 
set of force-free fields --- linear force free fields --- are easy to 
calculate (e.g.\ Nakagawa and Raadu 1972; Chiu and Hilton 1977; 
Alissandrakis, 1981; Gary, 1989). 
For these fields $\nabla\times {\bf B}=\alpha {\bf B}$ where 
$\alpha$ is a constant, the force-free parameter. Linear force-free models for 
coronal magnetic fields permit detailed descriptions of current paths in
the corona, although they have limitations as large scale models for 
solar fields (e.g.\ Sturrock, 1994).

In this paper the accuracies of simple circuit models for coronal
magnetic field energies are tested by application to linear force-free field
configurations. To our knowledge this is the first quantitative test of the
accuracy of the circuit model in the solar context. For simplicity, the 
comparison is limited to field configurations representing one and two 
magnetic loops. The order of presentation is as follows. In \S\,2 the
method is outlined, including the adopted boundary conditions (\S\,2.1),
and the details of the linear force-free (\S\,2.2) and circuit (\S\,2.3)
models. In \S\,3 the results of the calculations for one loop (\S\,3.1)
and two loops (\S\,3.2) are presented. Finally in \S\,4 the results are
discussed.

\section{Method}

\subsection{Boundary conditions}

\noindent
The plane $z=0$ is taken to represent the photosphere, and the half
space above this plane ($z>0$) represents the coronal volume. We consider 
simple boundary field configurations suggestive of one coronal loop:
\begin{equation}\label{eq:bc1}
B_z(x,y,0)=
  B_1\exp\left[-\frac{({\bf x}-{\bf x}_{1+})^2}{2\sigma_1^2}\right]
  -B_1\exp\left[-\frac{({\bf x}-{\bf x}_{1-})^2}{2\sigma_1^2}\right],
\end{equation}
and two coronal loops:
\begin{eqnarray} \label{eq:bc2}
B_z(x,y,0) & = &
  B_1\exp\left[-\frac{({\bf x}-{\bf x}_{1+})^2}{2\sigma_1^2}\right]
  -B_1\exp\left[-\frac{({\bf x}-{\bf x}_{1-})^2}{2\sigma_1^2}\right] \nonumber \\
  &+& B_2\exp\left[-\frac{({\bf x}-{\bf x}_{2+})^2}{2\sigma_2^2}\right]
  -B_2\exp\left[-\frac{({\bf x}-{\bf x}_{2-})^2}{2\sigma_2^2}\right],
\end{eqnarray}
respectively, where ${\bf x}=x\hat{\bf x}+y\hat{\bf y}$. 
The positions ${\bf x}_{1\pm}$ and ${\bf x}_{2\pm}$ 
define the loop footpoints.

It is convenient to adopt non-dimensional units in which 
lengths are expressed in terms of a basic length $L$, 
and magnetic fields are expressed in terms of the peak vertical field 
$B_1$ at the base of loop one, e.g.\ 
\begin{equation}
x=L\overline{x}, \quad B_z=B_1\overline{B}_z, \quad {\rm etc.},
\end{equation}  
where bars indicate dimensionless versions of quantities. 

\subsection{Linear force-free model}

\noindent
A solution to the linear force-free equations is provided by
(e.g.\ Alissandrakis, 1981)
\begin{eqnarray}\label{eq:ss_sol}
b_x(u,v,\overline{z})&=&\frac{-i(uk-v\overline{\alpha})}{2\pi (u^2+v^2)}
  e^{-k\overline{z}}b_z^0(u,v),
  \nonumber \\
b_y(u,v,\overline{z})&=&\frac{-i(vk+u\overline{\alpha})}{2\pi (u^2+v^2)}
  e^{-k\overline{z}}b_z^0(u,v), 
  \nonumber \\
b_z(u,v,\overline{z})&=&e^{-k\overline{z}}b_z^0(u,v), 
\end{eqnarray} 
where $b_i(u,v,\overline{z})$ denotes the Fourier transform of 
$\overline{B}_i(\overline{x},\overline{y},\overline{z})$ 
in $\overline{x}$ and $\overline{y}$ ($u$ and $v$ are the wavenumbers 
corresponding to $\overline{x}$ and $\overline{y}$ respectively), 
$b_z^0(u,v)=b_z(u,v,0)$ 
is the Fourier transform of 
$\overline{B}_z(\overline{x},\overline{y},0)$, 
$\overline{\alpha}=\alpha L$ is the force-free parameter, and 
\begin{equation}
k=\left[ 4\pi^2 (u^2+v^2)-\overline{\alpha}^2\right]^{1/2}.
\end{equation}

Equation~(\ref{eq:ss_sol}) 
requires $(u^2+v^2)^{1/2} \geq |\overline{\alpha}|/(2\pi)$. 
Since large 
$u$ and $v$ correspond to variation on small spatial scales, this
solution is termed the ``small scale solution'' to the linear 
force-free field equations. Of course, inversion of this solution in
Fourier space involves integration over all $u$ and $v$. The small
scale solution only provides a solution matching given boundary 
conditions $\overline{B}_z(\overline{x},\overline{y},0)$ 
if $b_z^0(u,v)$ is identically zero for 
$(u^2+v^2)^{1/2} < |\overline{\alpha}|/(2\pi)$. 
In general this is not true, and
in particular it will not be true for the boundary 
conditions~(\ref{eq:bc1}) and~(\ref{eq:bc2}), for any given 
$\overline{\alpha }\neq 0$ [although $b_z^0(u,v)$ is expected to be small 
for small $u$ and $v$, and identically zero for $u=v=0$ provided the net
flux is zero]. In the following we effectively calculate
$b_z^0(u,v)$ for the given boundary conditions and then impose 
$b_z^0(u,v)=0$ for $(u^2+v^2)^{1/2} < |\overline{\alpha}|/(2\pi)$. 

The physical justification for considering only the small scale 
solution is that it provides a linear force-free field 
with finite energy.
Following Alissandrakis (1981) and Gary (1989), using Parseval's 
theorem the energy of the linear force-free field~(\ref{eq:ss_sol}) 
in the half space $\overline{z}>0$ is 
\begin{equation}\label{eq:E_F}
\overline{E}_F=\int_{-\infty}^{\infty}du\int_{-\infty}^{\infty}dv
  \frac{|b_z^0(u,v)|^2}{\left[4\pi^2 (u^2+v^2)-\overline{\alpha}^2 
  \right]^{1/2}}.
\end{equation}
Once again this expression requires the Fourier components 
$b_z^0(u,v)$ to be identically zero for $u$ and $v$ such that 
$(u^2+v^2)^{1/2}< |\overline{\alpha}|/(2\pi)$. The large scale components 
of the field have infinite energy, and in this sense the general linear 
force-free boundary value problem has no physical solution 
(Alissandrakis, 1981). In practice we use the 
Fast Fourier Transform (FFT) of the boundary field defined on an $N$ by 
$N$ grid over the region $0\leq \overline{x}\leq 1$, 
$0\leq \overline{y}\leq 1$. The smallest non-zero wavenumber that is
determined from these `observations' of the field (by analogy with 
magnetograph observations) is $u_{\rm min}=v_{\rm min}=1-1/N$. It 
follows that the energies determined from the discrete
counterpart of~(\ref{eq:E_F}) are finite for all $N$ provided 
$|\overline{\alpha}|< \overline{\alpha}_{\rm max}=2\pi$. 

Finally we note that the variation of~(\ref{eq:E_F}) with 
$\overline{\alpha}$ for small values
of the force-free parameter is made obvious by a binomial expansion:
\begin{eqnarray}\label{eq:binomial}
\overline{E}_F(\overline{\alpha})
  =\overline{E}_F(0)
  &+&\frac{\overline{\alpha}^2}{16\pi^3}\int\!\!\int 
    \frac{|b_z^0(u,v)|^2}{(u^2+v^2)^{3/2}}dudv \nonumber \\
  &+&\frac{3\overline{\alpha}^4}{256\pi^5}\int\!\!\int 
    \frac{|b_z^0(u,v)|^2}{(u^2+v^2)^{5/2}}dudv +\cdots .
\end{eqnarray}
Hence for small $\overline{\alpha}$ the non-potential component of the
field scales with $\overline{\alpha}^2$.

\subsection{Circuit model}

\noindent
The current in a loop in the circuit model is taken to be the
axial current contained within a radius $\overline{R}$ of the centre of a 
footpoint:
\begin{eqnarray}\label{eq:I_1}
\overline{I}_1
  &=&2\pi \int_{0}^{\overline{R}}\overline{J}_z\overline{r}\,d\overline{r} 
  \nonumber \\
  &\approx&2\pi\overline{\alpha} \overline{\sigma}_1^2
  \left( 1-e^{-\overline{R}^2/(2\overline{\sigma}_1^2)} \right),
\end{eqnarray}
using the boundary conditions~(\ref{eq:bc1}) or~(\ref{eq:bc2}), and 
assuming that the footpoints are well separated.
Similarly
\begin{equation}\label{eq:I_2}
\overline{I}_2
  \approx 2\pi\overline{\alpha} \beta \overline{\sigma}_2^2
  \left( 1-e^{-\overline{R}^2/(2\overline{\sigma}_2^2)} \right),
\end{equation}
where $\beta\equiv B_2/B_1$, and $\overline{J}_z=\mu_0 L J_z/B_1$. 

The energy of the single loop model in the half space 
$\overline{z}>0$ is then estimated using~(\ref{eq:L}):
\begin{equation}\label{eq:E_C1}
\overline{E}_C=\frac{1}{2}\overline{L}_1\overline{I}_1^2,
\end{equation}
where
\begin{equation}\label{eq:L_1}
\overline{L}_1=\overline{a}_1\overline{C}_1, \quad {\rm with}
\quad
\overline{C}_1=\ln (8\overline{a}_1/\overline{R})-7/4.
\end{equation}
 
Similarly the energy of the two loop system can be expressed as 
\begin{equation}\label{eq:E_C2}
\overline{E}_C=\frac{1}{2}\overline{L}_1\overline{I}_1^2
+\frac{1}{2}\overline{L}_2\overline{I}_2^2
+\overline{M}_{12}\overline{I}_1\overline{I}_2,
\end{equation}
where $\overline{M}_{12}$ is the mutual inductance. For the general
case (loop major radii $\overline{a}_1$, $\overline{a}_2$, minor radii 
$\overline{r}_1$, $\overline{r}_2$, centres separated by 
$\overline{d}_{12}$, and axes oriented at an angle $\theta_{12}$) 
the non-dimensional version of the interpolating formula
Equation~(\ref{eq:M_don}) from Melrose (1997) may be used: 
\begin{equation}\label{eq:M_12}
\overline{M}_{12}=\frac{8\overline{a}_1^2\overline{a}_2^2
  (\overline{C}_1\overline{C}_2)^{1/2}\cos\theta_{12}}
  {\left[(\overline{a}_1+\overline{a}_1)^2+\overline{d}_{12}^2\right]^{3/2}}.
\end{equation}
For the more restricted case of line-current loops with their axes 
aligned ($\theta_{12}=0$) the non-dimensional version of the exact 
Equation~(\ref{eq:stratton}) may be used: 
\begin{equation}\label{eq:stratton_nd}
\overline{M}_{12}=\mu_0(\overline{a}_1\overline{a}_2)^{1/2}
\left[\left(\frac{2}{\overline{k}_{12}}-\overline{k}_{12}\right)
K(\overline{k}_{12}) -\frac{2}{\overline{k}_{12}}E(\overline{k}_{12})
\right],
\end{equation} 
where
\begin{equation}
\quad \overline{k}_{12}^2=\frac{4\overline{a}_1\overline{a}_2}
{(\overline{a}_1+\overline{a}_2)^2+\overline{d}_{12}^2}.
\end{equation}
 
\section{Results}

\subsection{One loop}

\noindent
Figure~1 illustrates a single loop configuration, for the boundary conditions
$\overline{{\bf x}}_{1+}=(0.5,0.1)$, $\overline{{\bf x}}_{1-}=(0.5,-0.1)$,
$\overline{\sigma}_1=0.05$, and $\overline{\alpha}=1$. 
(These values are adopted as nominal
values.) To construct this figure, the linear force-free field corresponding
to~(\ref{eq:ss_sol}) was calculated [using the FFT of Equation~(\ref{eq:bc1})]
on a $128\times 128\times 64$ grid corresponding to the region 
$0\leq \overline{x}\leq 1$, $0\leq \overline{y}\leq 1$, 
$0\leq \overline{z}\leq 1$. 
Field lines were traced for points at a radius 
$\overline{\sigma}_1/2=0.025$ from the positive footpoint.

\begin{figure}[ht]
\vspace{0.5cm}
\centerline{\epsfig{file=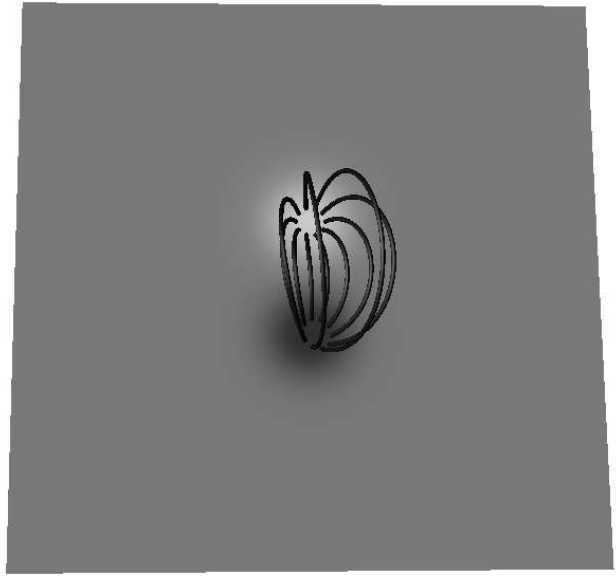,width=20pc}}
\caption{\label{fig:f1}A single loop configuration.} 
\end{figure}   

The free energy $\overline{E}_F(\overline{\alpha})-\overline{E}_F(0)$ 
of the field can be estimated by taking the FFT of the boundary field 
with $N$ gridpoints and then evaluating the integral~(\ref{eq:E_F}) 
using the trapezoidal rule. We label this estimate $W_N$. 
Figure~2 shows how the result depends on $N$, for the nominal choices 
of $\overline{{\bf x}}_{1\pm}$, $\overline{\sigma}_1$, and 
$\overline{\alpha}$. The upper panel plots $W_N$ versus $N$, for
$N=32,64,128,...,4096$. The lower panel plots $W_N/W_{N/2}-1$ versus 
$N$ as a log-log plot. A condition for the sequence of values of 
$W_N$ to converge is that $W_N/W_{N/2}$ approaches unity as 
$N\rightarrow\infty$. The lower panel suggests that this is occurring,
and moreover $W_N/W_{N/2}-1\sim N^{-1}$.
For the calculations in the paper we have used $N=512$. The estimates 
of $W_N$ for $N=512$ and $N=4096$ differ by about $0.5\%$, which suggests 
that $N=512$ is sufficiently accurate for present purposes.

\begin{figure}[ht]
\vspace{0.5cm}
\centerline{\epsfig{file=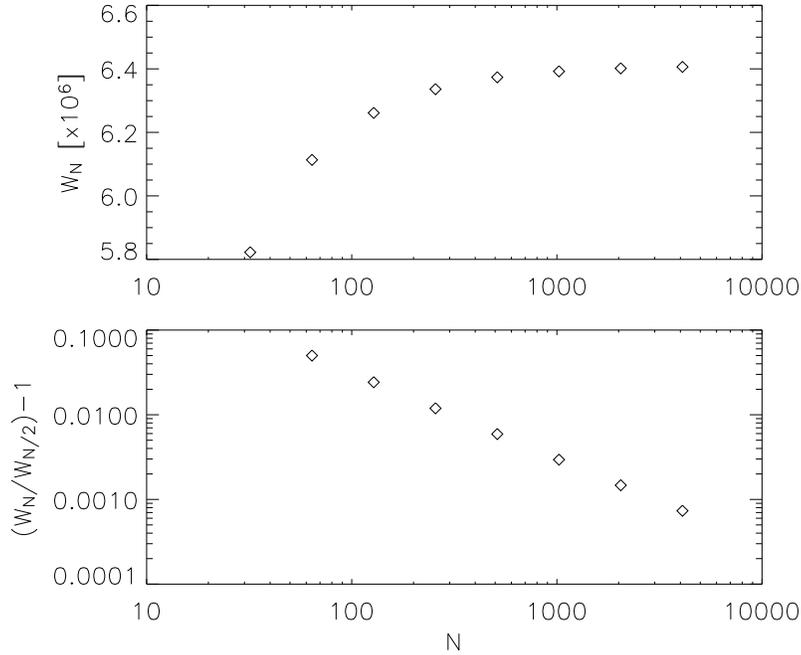,width=25pc}}
\caption{\label{fig:f2}Upper: Variation of the free energy estimate
$W_N$ with $N$, for the single loop configuration. Lower: 
Convergence of the estimates.} 
\end{figure}   

The first question to be addressed is the variation of the energy
estimates with $\overline{I}_1$. To vary $\overline{I}_1$ we have varied
$\overline{\alpha}$, while keeping $\overline{\sigma}_1$ fixed [see
Equation~(\ref{eq:I_1})]. We have
also chosen $\overline{R}=\overline{\sigma}_1$, and used the nominal
values of $\overline{{\bf x}}_{1\pm}$ and $\overline{\sigma}_1$. The
field estimate of the energy $W_{512}$ and the circuit estimate $E_C$ were 
calculated for 50 values of $\overline{\alpha}$ between 
$0.001\overline{\alpha}_{\rm max}$ and $0.99\overline{\alpha}_{\rm max}$,
with the values being equally spaced logarithmically. The results are
plotted in Figure~3. The diamonds show the values of $W_{512}$ and the 
crosses show the values of $E_C$. The vertical dashed line is 
$\overline{\alpha}_{\rm max}$ and the horizontal dashed line is the
estimate of $\overline{E}_F(0)$, the energy of the potential component
of the field. The energy estimates for the non-potential component of the
field and the circuit energy estimates are generally much smaller than the
energy of the potential component of the field. 
The energy estimates for the non-potential component of the field 
exhibit a $\overline{I}_1^2$ dependence ($\overline{\alpha}^2$ dependence) 
for most of the chosen range, but increase more rapidly for large 
$\overline{I}_1$ ($\overline{\alpha}$). This functional dependence is 
consistent with Equation~(\ref{eq:binomial}). The field and circuit
energy estimates are within a factor of two of one another for most
of the chosen range, but are out by more than a factor of 10 close
to the maximum value of $\overline{\alpha}$. There is a simple physical
interpretation for the observed dependence of the field energy on 
$\overline{I}$. For small values of the current the field associated
with the current is small, and the magnetic field is dominated by the
potential component. The current essentially follows the fieldlines of 
the potential field, i.e.\ there is a fixed geometrical path for the 
current. In this case the self-inductance defined by 
Equation~(\ref{eq:general_L}) is independent of the
current. For larger values of the current this is no longer the case: the
field due to the current becomes comparable to the potential component of
the field, and the path that the current takes then depends on the current.
In particular, larger currents will tend to produce longer current paths,
so the effective self-inductance will increase with current.
In other words Equation~(\ref{eq:general_L}) increases with 
current. 
 
\begin{figure}[ht]
\vspace{0.5cm}
\centerline{\epsfig{file=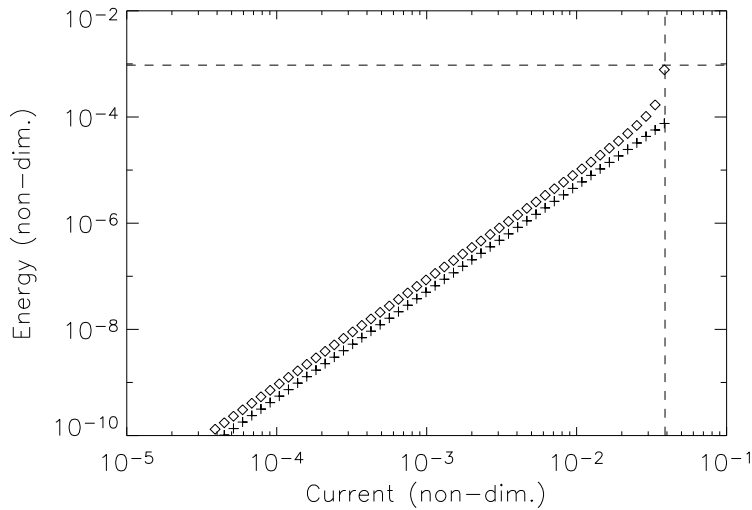,width=25pc}}
\caption{\label{fig:f3}Current dependence of energy estimates for 
the single loop configuration.} 
\end{figure}   

Another question is the dependence of energy on the geometry of the
loop. The major radius $\overline{a}$ was varied between 0.05 and
0.25 for the nominal choices of the other parameters, and Figure~4 shows
the results. The energy estimates $W_{512}$ were divided 
by $\frac{1}{2}\overline{I}_1^2$ [see Equation~(\ref{eq:L})] to 
produce effective self-inductances, which are shown by diamonds. 
Results are shown for $\overline{\alpha}=1.0$ and $\overline{\alpha}=2.0$
--- the effective self-inductances for the larger value of 
$\overline{\alpha}$ are slightly larger, for all values of major radius.
The self-inductances $\overline{L}_1$ are indicated by crosses. 
This figure shows that the non-potential component
of the field energy varies with major radius in a manner similar to
that expected from the circuit model for the chosen range of values.
The inferred self-inductance increases monotonically, as expected 
on the basis of the increase in area under the loop. However the precise
functional dependence of the inferred self-inductance is somewhat 
different from that expected from the simple model $\overline{L}_1$. 
The similar
results for the different values of $\overline{\alpha}$ show
that the inferred self-inductance is almost independent of the value
of the current, as expected on the basis of the argument given above.

\begin{figure}[ht]
\vspace{0.5cm}
\centerline{\epsfig{file=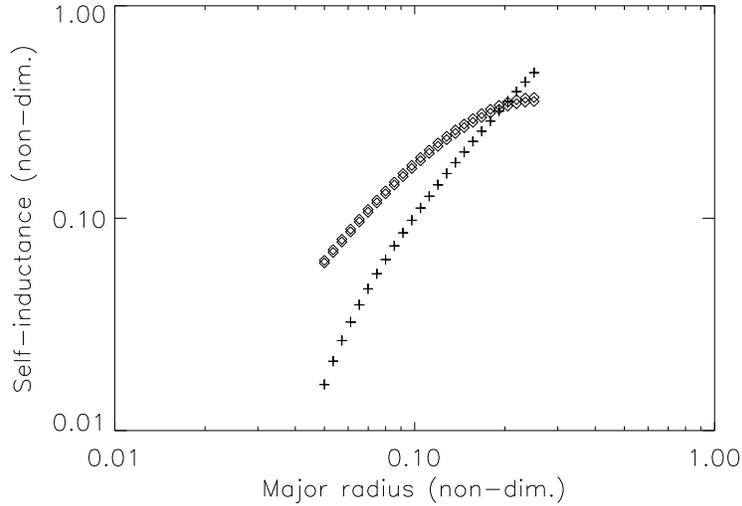,width=25pc}}
\caption{\label{fig:f4}Major radius dependence of self-inductance
for the single loop configuration.} 
\end{figure}   

The minor radius dependence of energy was also examined. Figure~5
shows the effective self-inductance associated with the non-potential
component of the magnetic field (diamonds) for $\overline{\sigma}_1$ between
0.01 and 0.1, and for the nominal values of the other parameters.
The values of $\overline{L}_1$ are also plotted (crosses). This figure
shows that the variation of the magnetic field energy with 
$\overline{\sigma}_1$ is qualitatively consistent with the circuit 
model. The effective self-inductance decreases with 
$\overline{\sigma}_1$, as expected from the decrease in area within
within the loop as the minor radius of the loop increases.

\begin{figure}[ht]
\vspace{0.5cm}
\centerline{\epsfig{file=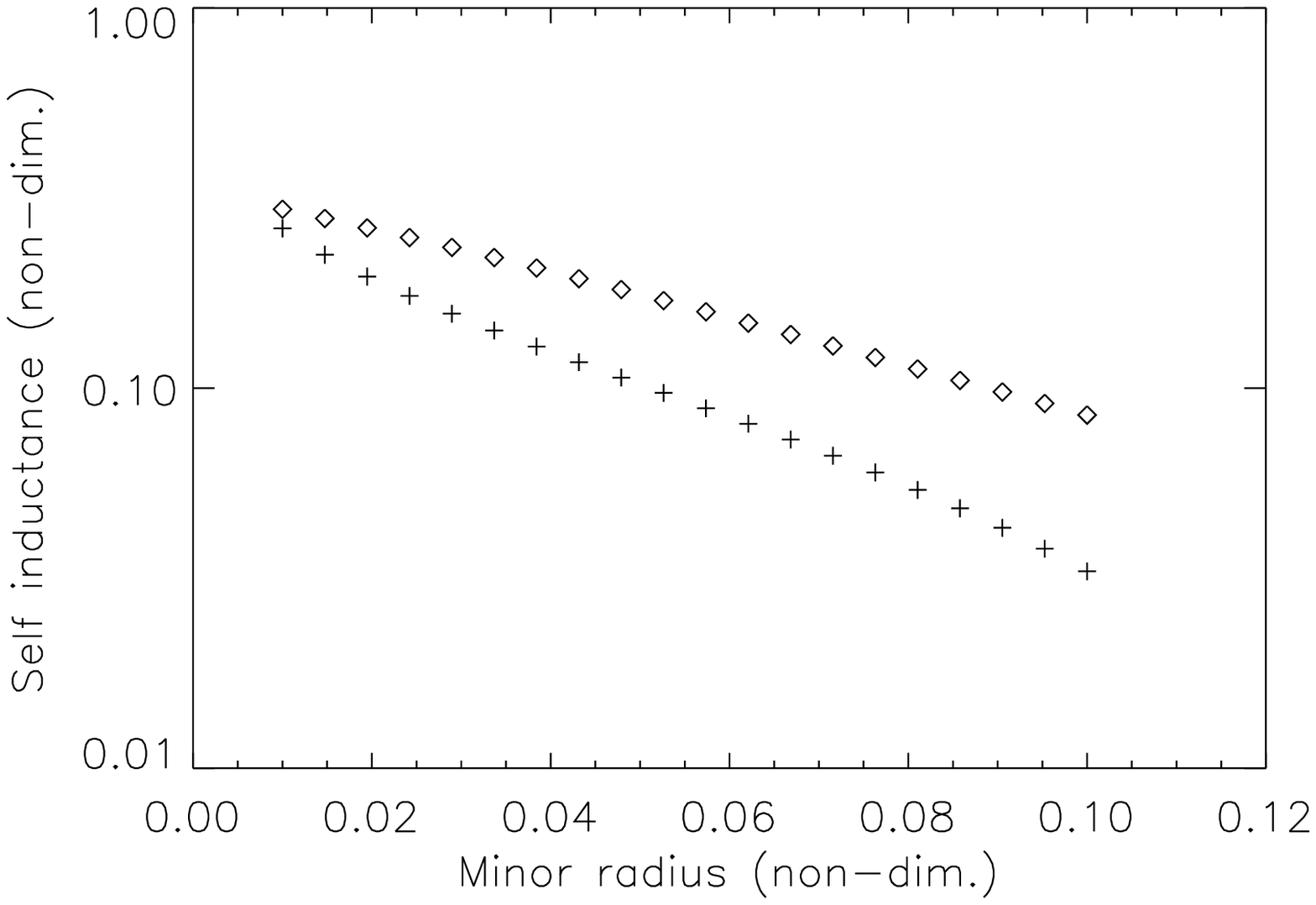,width=25pc}}
\caption{\label{fig:f5}Minor radius dependence of self-inductance
for the single loop configuration.} 
\end{figure}   

\subsection{Two loops}

\noindent
Figure~6 illustrates a particular two-loop configuration, for the 
boundary conditions $\overline{{\bf x}}_{1+}=(0.25,0.575)$,
$\overline{{\bf x}}_{1-}=(0.25,0.425)$,
$\overline{{\bf x}}_{2+}=(0.825,0.5)$,
$\overline{{\bf x}}_{2-}=(0.675,0.5)$, 
$\overline{\sigma}_1=\overline{\sigma}_2=0.025$, $\beta=1$, 
and $\overline{\alpha}=1$. 
These choices correspond to loops with
major radii $\overline{a}_1=\overline{a}_2=0.075$, with a distance 
$\overline{d}=0.5$ between their centres, and with axes at right 
angles to one another ($\theta_{12}=\pi/2$).

\begin{figure}[ht]
\vspace{0.5cm}
\centerline{\epsfig{file=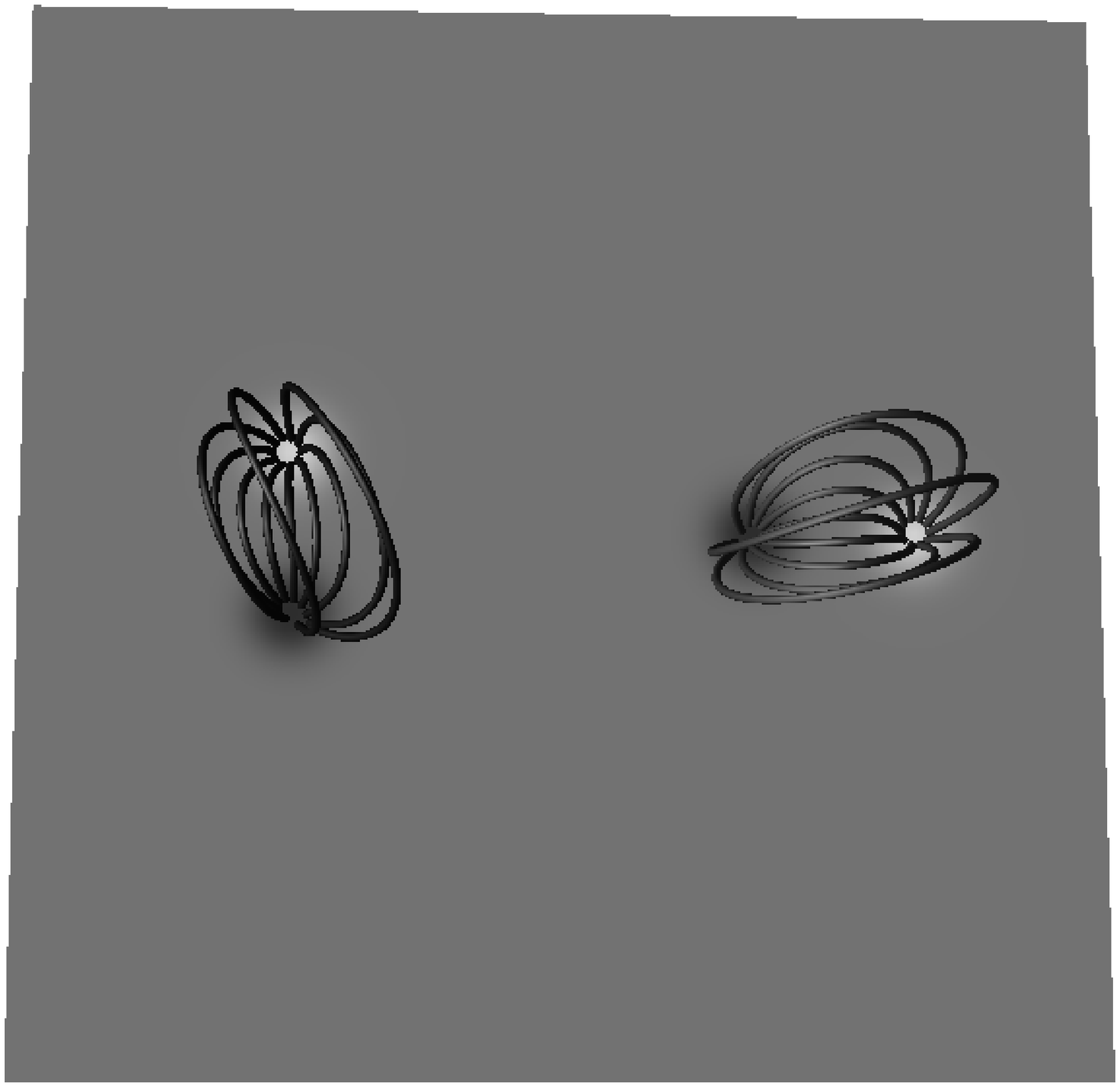,width=20pc}}
\caption{\label{fig:f6}A two-loop configuration with 
$\theta_{12}=\pi/2$.} 
\end{figure} 

There are many energy dependences which could be investigated for 
two-loop configurations. However, from the point of view of the circuit
model the interest is with the energy associated with the interaction
between the loops, described by the mutual inductance. The 
interaction energy is determined in the following way. The field 
energy $W_{512}$ for the two loops is calculated. Then the field energy
$W_{512,1}$ for loop one is calculated (the free energy 
of the small scale linear force-free field with the boundary conditions 
of loop one). Similarly the field energy $W_{512,2}$
for loop two is calculated. The energy $W_{512}-W_{512,1}-W_{512,2}$
is the interaction energy. Dividing this quantity by 
$\overline{I}_1\overline{I}_2$ defines an effective mutual inductance
$\overline{M}_{12}^{\rm eff}$, which may be compared with circuit 
estimates.

First we consider a comparison with the Stratton 
(1941) formula~(\ref{eq:stratton_nd}), since it is an exact expression
for mutual inductance (albeit for line currents). Recall that it is
applicable only to the case of loops with aligned axes 
($\theta_{12}=0$). The dependence of the mutual inductance on the major
radius of loop two was investigated for loop parameters 
$\overline{a}_1=0.075$, $\overline{d}=0.5$,
$\overline{\sigma}_1=\overline{\sigma}_2=0.025$, $\theta_{12}=0$, $\beta=1$, 
$\overline{\alpha}=1$, with the major radius of loop two taking 20
values in the range $0.02$ to $0.2$. Figure~7 shows the results. The
points marked by diamonds are the mutual inductances calculated for the
field following the procedure outlined above. The points marked with
crosses are the mutual inductances given by the 
formula~(\ref{eq:stratton_nd}). The inferred mutual inductances are
qualitatively consistent with the circuit values, and a similar
functional dependence is observed.

\begin{figure}[ht]
\vspace{0.5cm}
\centerline{\epsfig{file=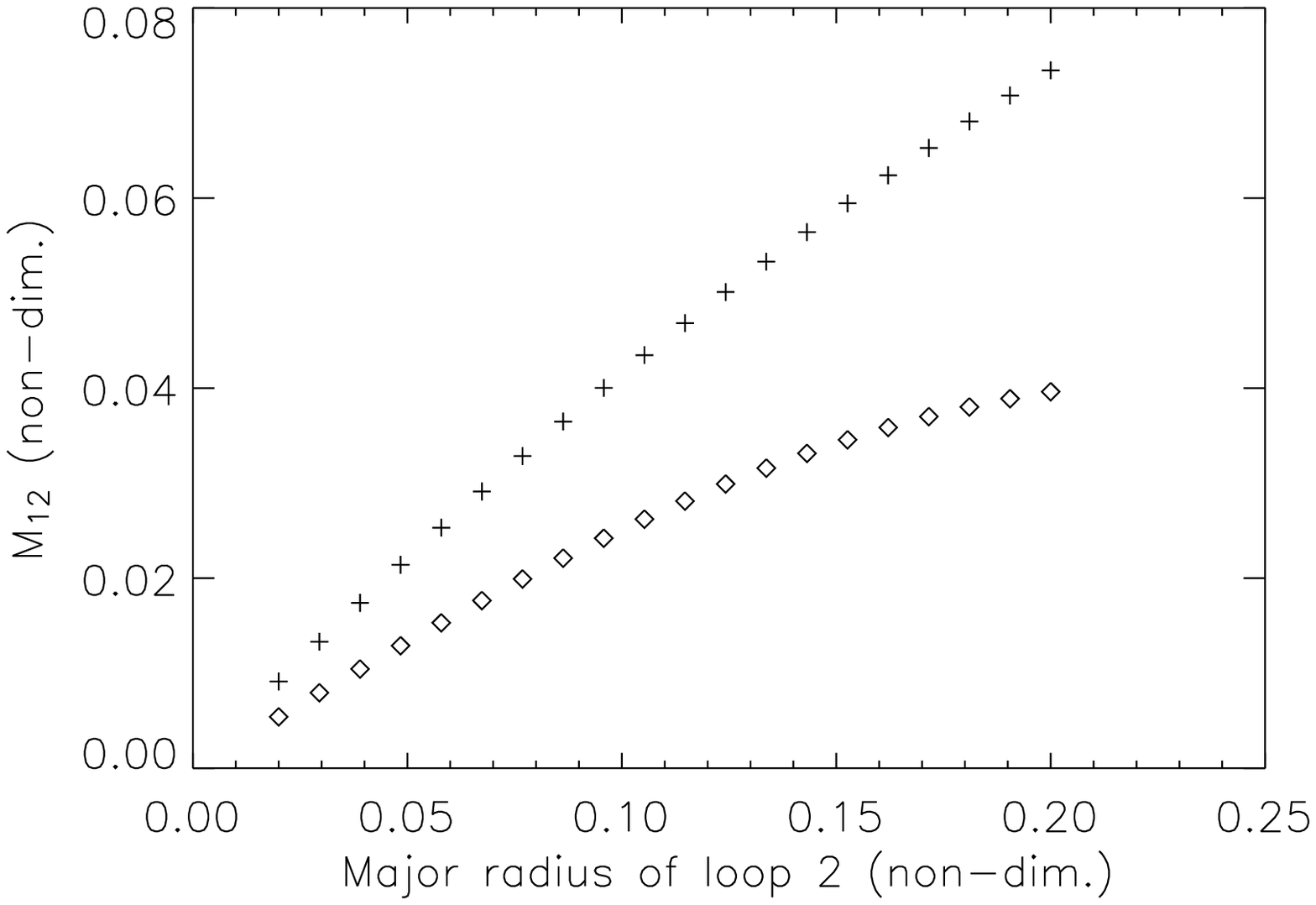,width=25pc}}
\caption{\label{fig:f7}Variation of effective inductance with 
major radius of loop 2 for the two-loop configuration, and 
comparison with the Stratton (1941) formula.} 
\end{figure} 

Next we examine the Melrose (1997) interpolation formula~(\ref{eq:M_12}), 
which is more general than the Stratton (1941) formula, but is 
approximate. We considered a two-loop configuration with fixed parameters 
but with different values of $\theta_{12}$, corresponding to a rotation of 
loop two with respect to loop one. Specifically the values
$\overline{a}_1=\overline{a}_2=0.075$, $\overline{d}=0.5$,
$\overline{\sigma}_1=\overline{\sigma}_2=0.025$, $\beta=1$, 
and $\overline{\alpha}=1$ were chosen, together with 
21 values of $\theta_{12}$ in the range 
$0\leq \theta_{12}\leq \pi$. Figure~8 shows
the results. The crosses are the circuit values. Following 
Equation~(\ref{eq:M_12}) the circuit model exhibits a simple cosine 
dependence on $\theta_{12}$, and in particular is zero for 
$\theta_{12}=\pi/2$, because perpendicular loops have no 
flux linkage. The diamonds are the values of 
$\overline{M}_{12}^{\rm eff}$ 
determined following the procedure 
outlined above. The observed functional dependence is essentially 
cosine, and indeed the effective mutual inductance is very close to 
zero for $\theta_{12}=\pi/2$. However, the values of 
$\overline{M}_{12}^{\rm eff}$ are considerably larger than the values
given by the Melrose formula: the ratio of the values is close to eight 
over the range of $\theta_{12}$. This figure also shows the mutual
inductance obtained using the Stratton (1941) formula for the special
case $\theta_{12}=0$ as an asterisk. The Stratton value is slightly 
larger than $\overline{M}_{12}^{\rm eff}$ for $\theta_{12}=0$, and 
is considerably larger than the Melrose value. 
It is notable that the Melrose value differs
from the Stratton value by more than an order of magnitude in this case.
 
\begin{figure}[ht]
\vspace{0.5cm}
\centerline{\epsfig{file=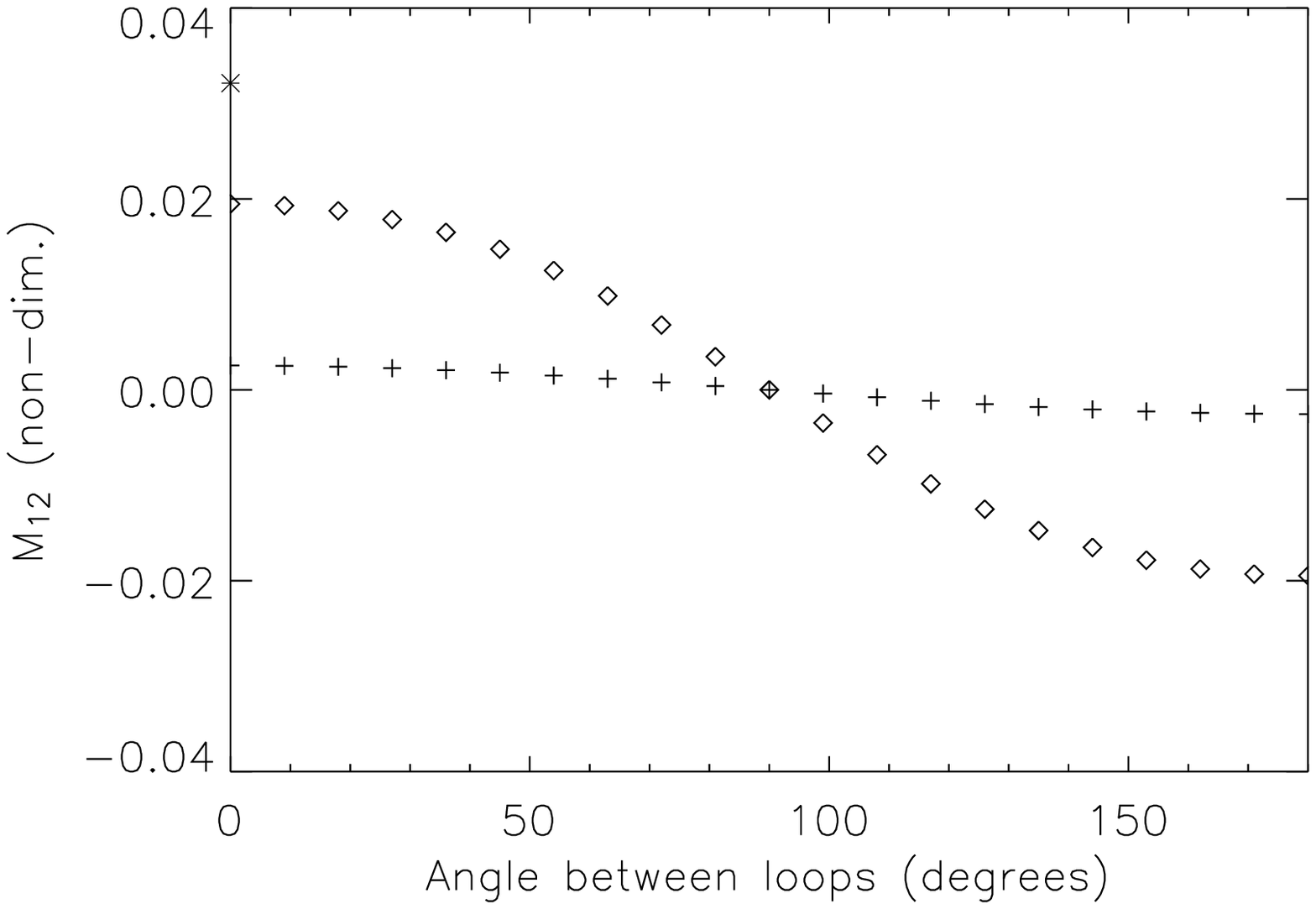,width=25pc}}
\caption{\label{fig:f8}Variation of effective inductance with angle
between loops for the two-loop configuration, and comparison with
the Melrose (1997) formula.} 
\end{figure} 

\section{Discussion and conclusions}

\noindent
In this paper calculated energies of non-potential components of linear
force-free fields are compared with simple circuit model estimates,
for boundary conditions suggestive of one and two magnetic loops.
The motivation is to test the circuit models, which are often invoked to 
estimate coronal magnetic field energies. In particular circuit models
have been used to describe flare-related energy changes. In the Melrose 
(1997) model, preferred magnetic configurations for flaring are identified
on the basis of geometrical dependences of circuit energy estimates. 

In \S\,3.1 results for a single loop configuration are presented. An
important result is the $\overline{\alpha}$-dependence of the energy of
the non-potential component of the field. For a range of small values of 
the force-free parameter $\overline{\alpha}$ the free energy is found 
to be proportional to $\overline{\alpha}^2$, or
equivalently to be proportional to the square of the current in the loop,
consistent with the circuit model.
This result is explained mathematically in terms of the 
$\overline{\alpha}^2$ dependence of the free energy appearing in 
Equation~(\ref{eq:binomial}) for small $\overline{\alpha}$.  
For larger values of current (or $\overline{\alpha}$) the free energy 
increases more rapidly than the circuit prediction.
This result is also expected for the two-loop configuration, since 
Equation~(\ref{eq:binomial}) applies for any boundary conditions. A simple
physical interpretation of this result is as follows. For 
small values of $\overline{\alpha}$ the non-potential component of the 
field is small compared to the potential component, and the current 
essentially follows the fieldlines of 
the potential field. Hence as $\overline{\alpha}$ is varied the geometry 
of the current does not change and the inductances [which are geometrical 
quantities --- see Equations~(\ref{eq:general_L}) 
and~(\ref{eq:general_M})] are constant. 
This situation corresponds to the free energy 
scaling with the square of the current (or $\overline{\alpha}^2$). For 
larger values of $\overline{\alpha}$ the non-potential component of the 
field becomes comparable to the potential component, and influences the 
path of the current. In this case the geometry [and hence the 
inductances as defined by Equations~(\ref{eq:general_L}) 
and~(\ref{eq:general_M})] depend on $\overline{\alpha}$, which 
corresponds to the free 
energy scaling with a higher power of $\overline{\alpha}$. Although 
demonstrated here for a class of linear force-free fields, this effect is
expected to be quite general and to limit the accuracy of simple 
circuit models
in application to any force-free field for large values of current.

In \S\,3.1 the dependence of the free energy of the field on the 
geometry of the loop is also investigated (for a modest value of 
$\overline{\alpha}$). The functional dependences on minor radius and 
major radius are found to be qualitatively consistent with the circuit 
model.

Section \S\,3.2 presents the results for the two-loop configuration.
The interaction energy of the two loops is obtained by calculating the
free energy of the loops together and then subtracting
off the individual free energies of each loop. This procedure 
is used to calculate an effective mutual inductance 
$\overline{M}_{12}^{\rm eff}$ for the field. This was then compared with
simple circuit model formulae, namely an exact expression for the
mutual inductance of two axially-aligned line-current loops due to Stratton 
(1941), and an approximate expression for the mutual inductance of oblique 
loops with finite cross section due to Melrose (1997). The Stratton formula
is found to qualitatively reproduce the observed variation of
$\overline{M}_{12}^{\rm eff}$ with the major radius of one loop. 
The Melrose formula is compared with $\overline{M}_{12}^{\rm eff}$ as
one loop is rotated with respect to the other. The effective mutual
inductance of the field reproduces the cosine dependence in the Melrose
formula, and in particular the inferred mutual inductance is close 
to zero for the case of perpendicular loops. However, the calculated 
values of $\overline{M}_{12}^{\rm eff}$ are almost an order of magnitude
larger than the Melrose formula values. Comparison with the Stratton
formula for the case of parallel loops confirms that, for the chosen
parameters, the Melrose formula underestimates the actual interaction
energy. The Melrose expression is an interpolation between known results,
and so the discrepancy can be attributed to the approximate nature of
the formula.

In this paper linear force-free fields have been used because they are
straightforward to calculate.
It is likely that real coronal magnetic fields involve currents that
are spatially highly concentrated, and hence require a nonlinear 
force-free model (or a description involving non-zero pressure forces). 
In this case circuit models may provide quite accurate energy 
estimates because the configuration closely resembles a set of isolated 
circuits. However, it is also possible that in the nonlinear case the
current path sensitively depends on the value of the current, in particular
for large currents, and this may limit the accuracy of simple circuit 
models, as discussed above.

Another point to note is that circuit models are really only useful 
if there are a small number of circuits which are easily identified. 
In this paper we have considered widely separated magnetic structures, 
which as a result have simple connectivity. In general coronal fields 
have a complicated topology, and it may be difficult to identify coronal
current paths.

Circuit models appear to provide reasonably accurate free energy 
estimates for coronal magnetic field configurations for a range of values
of the current, and also qualitatively describe various functional 
dependences of the free energy (the dependence on loop major radii, 
the dependence on the angle between two loops, etc.). Simple circuit
models are less accurate for very large values of the current, when the
current path depends on the value of the current. Despite this reservation,
the results of this paper confirm the general utility of circuit models, 
in particular in their application to the flare phenomenon. 

\section*{Acknowledgements}

\noindent 
M.S.W. acknowledges the support of an Australian Research Council 
QEII Fellowship.

\end{document}